\documentclass[runningheads,a4paper]{llncs}
\usepackage[english]{babel}
\usepackage[ruled,vlined,nosemicolon]{algorithm2e}
\usepackage{float}
\usepackage{tikz}
\usepackage{paralist}
\usepackage[inline]{enumitem}
\usepackage{amssymb,amstext,amsmath}
\usepackage{mathtools}
\usepackage{cleveref}

\usetikzlibrary{hobby,calc}

\input{define.orgtex}

\begin{document}

\title{Acacia-Bonsai: A Modern Implementation of Downset-Based LTL Realizability}
\titlerunning{Acacia-Bonsai: Downset-Based LTL Realizability}

\author{Micha\"el Cadilhac\inst{1}
\and Guillermo A. P\'erez\inst{2}
}
\institute{DePaul University, USA\\
\email{michael@cadilhac.name} \and
University of Antwerp -- Flanders Make, Belgium\\
\email{guillermoalberto.perez@uantwerpen.be}
}
\authorrunning{M. Cadilhac \and G. A. P\'erez}

\maketitle

\begin{abstract}
  We describe our implementation of downset-manipulating algorithms used to solve the
  realizability problem for linear temporal logic (LTL). These algorithms were
  introduced by Filiot et al.~in the 2010s and implemented in the tools Acacia
  and Acacia+ in C and Python.  We identify degrees of freedom in the original
  algorithms and provide a complete rewriting of Acacia in C++20 articulated
  around genericity and leveraging modern techniques for better performances.
  These techniques include compile-time specialization of the algorithms, the
  use of SIMD registers to store vectors, and several preprocessing steps, some
  relying on efficient Binary Decision Diagram (BDD) libraries.  We also explore
  different data structures to store downsets.  The resulting tool is
  competitive against comparable modern tools.
  \keywords{LTL synthesis \and C++ \and downset \and antichains \and SIMD \and BDD}
\end{abstract}

\section{Introduction}

Nowadays, hardware and software systems are everywhere around us. One way to
ensure their correct functioning is to automatically synthesize them from a
formal specification.  This has two advantages over alternatives such as
testing and model checking: the design part of the program-development process
can be completely bypassed and the synthesized program is correct by
construction.

In this work we are interested in synthesizing \emph{reactive
systems}~\cite{hp84}. These maintain a continuous interaction with their
environment.  Examples of reactive systems include communication, network, and
multimedia protocols as well as operating systems.  For the specification, we
consider \emph{linear temporal logic} (LTL)~\cite{pnueli77}. LTL allows to
naturally specify time dependence among events that make up the formal
specification of a system. The popularity of LTL as a formal specification
language extends to, amongst others, AI~\cite{gv16,cm19,gnpw20}, hybrid
systems and control~\cite{bvpyb16}, software engineering~\cite{lpb15}, and
bio-informatics~\cite{abbdfhinprs17}.

The classical doubly-exponential-time synthesis algorithm can be decomposed
into three steps:
\begin{enumerate*}
  \item \emph{compile} the LTL formula into an automaton of exponential
    size~\cite{vw84},
  \item \emph{determinize} the automaton~\cite{safra88,piterman07} incurring a
    second exponential blowup,
  \item and determine the winner of a \emph{two-player zero-sum game} played
    on the latter automaton~\cite{pr89}.
\end{enumerate*}
Most alternative approaches focus on avoiding the determinization step of the
algorithm. This has motivated the development of so-called Safra-less
approaches, e.g.,~\cite{kpv06,eks16,ekrs17,tushy17}. Worth mentioning are the
on-the-fly game construction implemented in the Strix tool~\cite{msl18} and
the \emph{downset}-based (or ``antichain-based'') on-the-fly bounded
determinization described in~\cite{fjr09} and implemented in
Acacia+~\cite{bbfjr12}. Both avoid constructing the doubly-exponential
deterministic automaton.  Acacia+ was not ranked in recent editions of
SYNTCOMP~\cite{syntcomp17} (see \url{http://www.syntcomp.org/}) since it is no longer
maintained despite remaining one of the main references for new advancements
in the field (see, e.g.,~\cite{ffrt17,ztlpv17,apsec20,lms20,bltv20}).

\paragraph*{Contribution.}
We present the Acacia approach to solving the problem at hand and propose a
new implementation that allows for a variety of optimization steps.  For now,
we have focused on \emph{(B\"uchi automata) realizability}, i.e., the decision
problem which takes as input an automaton compiled from the LTL formula and
asks whether a controller satisfying it exists. In our tool, we compile the
input LTL formula into an automaton using Spot~\cite{duret.16.atva2}.  We
entirely specialize our presentation on the technical problem at hand and
strive to distillate the algorithmic essence of the Acacia approach in that
context. The main algorithm is presented in Section~\ref{sec:algo} and the
different implementation options are listed in Section~\ref{sec:implem}.
Benchmarks are included in Section~\ref{sec:benchmarks}.

\section{Preliminaries}

Throughout this abstract, we assume the existence of two alphabets, \(I\) and
\(O\); although these stand for input and output, the actual definitions of these
two terms is slightly more complex: An \emph{input} (\resp \emph{output}) is a
boolean combination of symbols of \(I\) (\resp $O$) and it is \emph{pure} if it is
a \emph{conjunction} in which \emph{all} the symbols in \(I\) (\resp $O$) appear;
e.g., with \(I = \{i_1, i_2\}\), the expressions \(\top\) (true), \(\bot\) (false), and
\((i_1 \lor i_2)\) are inputs, and \((i_1 \land \neg i_2)\) is a pure input. Similarly,
an \emph{IO} is a boolean combination of symbols of \(I \cup O\), and it is
\emph{pure} if it is a conjunction in which all the symbols in \(I \cup O\) appear.
We use \(i, j\) to denote inputs and \(x, y\) for IOs.  Two IOs \(x\) and \(y\) are
\emph{compatible} if \(x \land y \neq \bot\).

A \emph{Büchi automaton} \cA is a tuple \((Q, q_0, \delta, B)\) with \(Q\) a set of
states, \(q_0\) the initial state, \(\delta\) the transition relation that uses IOs as
labels, and \(B \subseteq Q\) the set of Büchi states.  The actual semantics of this
automaton will not be relevant to our exposition, we simply note that these
automata are usually defined to recognize infinite sequences of symbols from
\(I \cup O\).  We assume, throughout this paper, the existence of some automaton
\(\cA\).

We will be interested in valuations of the states of \(\cA\) that indicate some
sort of progress towards reaching Büchi states---again, we do not go into details
here.  We will simply speak of \emph{vectors over \cA} for elements in
\(\bbZ^Q\), mapping states to integers.  We will write \(\vec{v}\) for such vectors,
and \(v_q\) for its value for state \(q\).  In practice, these vectors will range
into a finite subset of \(\bbZ\), with \(-1\) as an implicit minimum value (meaning
that \((-1) - 1\) is still \(-1\)) and an upper bound provided by the problem.

For a vector \(\vec{v}\) over \(\cA\) and an IO \(x\), we define a function that takes
one step back in the automaton, decreasing components that have seen Büchi
states.  Write \(\chi_B(q)\) for the function mapping a state \(q\) to \(1\) if
\(q \in B\), and \(0\) otherwise.  We then define \(\bwd(\vec{v}, x)\) as the vector
over \cA that maps each state \(p \in Q\) to:
\[\min_{\substack{(p, y, q) \in \delta\\ x \text{ compatible with } y}} \left(v_q -
  \chi_B(q)\right)\enspace,\]
and we generalize this to sets: \(\bwd(S, x) = \{\bwd(\vec{v}, x) \mid \vec{v}
\in S\}\).
For a set \(S\) of vectors over \cA and a (possibly nonpure) input \(i\), define:
\[\cpre_i(S) = S \cap \bigcup_{\substack{x \text{ pure IO}\\x \text{ compatible with } i}} \bwd(S, x)\enspace.\]

It can be proved that iterating \cpre with any possible pure input stabilizes to
a fixed point that is independent from the order in which the inputs are
selected.  We define \(\cpre^*(S)\) to be that set.

All the sets that we manipulate will be \emph{downsets}: we say that a vector
\(\vec{u}\) dominates another vector \(\vec{v}\) if for all \(q \in Q\),
\(u_q \geq v_q\), and we say that a set is a downset if \(\vec{u} \in S\) and
\(\vec{u}\) dominates \(\vec{v}\) implies that \(\vec{v} \in S\).  This allows to
implement these sets by keeping only dominating elements, which form, as they
are pairwise nondominating, an \emph{antichain}.  In practice, it may be
interesting to keep more elements than just the dominating ones or even to keep
all of the elements to avoid the cost of computing domination.

Finally, we define \(\safe_k\) as the smallest downset containing the all-\(k\)
vector.
We are now equipped to define the computational problem we focus on:\\[1em]
\textbf{BackwardRealizability}
\begin{compactitem}
\item \textbf{Given:} A Büchi automaton \cA and an integer \(k > 0\),
\item \textbf{Question:} Is there a \(\vec{v} \in \cpre^*(\safe_k)\) with \(v_{q_0} \geq
  0\)?
\end{compactitem}
\vspace{1em}

We note, for completeness, that (for sufficiently large values of $k$) this
problem is equivalent to deciding the realizability problem associated with
\cA: the question has a positive answer iff the \emph{output player} wins the
Gale-Stewart game with payoff set the complement of the language of \(\cA\).

\section{Realizability algorithm}

The problem 
admits a natural algorithmic solution: start with the initial
set, pick an input \(i\), apply \(\cpre_i\) on the set, and iterate until all inputs
induce no change to the set, then check whether this set contains a vector that
maps \(q_0\) to \(0\).  We first introduce some degrees of freedom in this approach,
then present a slight twist on that solution that will serve as a canvas for the
different optimizations.

\subsection{Boolean states}

This opportunity for optimization was identified in \cite{bohy14} and
implemented in Acacia+, we simply introduce it in a more general setting and
succinctly present the original idea when we mention how it
can be implemented in
Section~\ref{sec:implem-bool}.  We start with an example.  Consider the
Büchi automaton:

\pgfmathsetseed{41}
\tikzset{
 pics/blob/.style={
   code={
   \draw[use Hobby shortcut, fill, closed] (0,0) +($(0:1+4*rnd)$)
       \foreach \a in {60,120,...,350} {  .. +($(\a: 1+4*rnd)$) };
   }
}}

\begin{center}
  \begin{smallautomaton}
    \node[smallstate,initial] (q0) {\(q_0\)};
    \node[smallstate, right of=q0] (q1) {\(q_1\)};
    \path pic[right=1.1cm of q1,smallstate,scale=0.15,dashed] {blob};
    \node[right of=q1] (r) {};

    \path[->] (q0) edge node {\(\top\)} (q1) (q1) edge node {\(\top\)} (r);
  \end{smallautomaton}
\end{center}

Recall that we are interested in, after \cpre has stabilized, whether the
initial state can carry a nonnegative value.  In that sense, the crucial
information associated with \(q_0\) is boolean in nature: is its value positive or
\(-1\)?  Even further, this same remark can be applied to \(q_1\) since \(q_1\) being
valued \(6\) or \(7\) is not important to the valuation of \(q_0\).  Hence the set of
states may be partitioned into integer-valued states and boolean-valued ones.
Naturally, detecting which states can be made boolean comes at a
cost 
and not doing it is a valid option.

\subsection{Actions}

For each IO \(x\), we will have to compute \(\bwd(\vec{v}, x)\) oftentimes.  This
requires to refer to the underlying Büchi automaton and checking for each
transition therein whether \(x\) is compatible with the condition.  It may be
preferable to precompute, for each \(x\), what are the relevant pairs \((p, q)\) for
which \(x\) can go from \(p\) to \(q\).  We call the set of such pairs the
\emph{io-action} of \(x\) and denote it \(\ioact(x)\); in symbols:
\[\ioact(x) = \{(p, q) \mid (\exists (p, y, q) \in \delta)[x \text{ is compatible with }
y]\}\enspace.\]

Further, as we will be computing \(\cpre_i(S)\) for inputs \(i\), we abstract in a
similar way the information required for this computation.  We use the term
\emph{input-action} for the set of io-actions of IOs compatible with \(i\) and denote
it \(\iact(i)\); in symbols:
\[\iact(i) = \{\ioact(x) \mid x \text{ is an IO compatible with } i\}\enspace.\]

In other words, actions contain exactly the information necessary to compute
\(\cpre\).  Note that from an implementation point of view, we do not require that
the actions be precomputed.  Indeed, when iterating through pairs
\((p, q) \in \ioact(x)\), the underlying implementation can choose to go back to the
automaton.

\subsection{Sufficient inputs}\label{sec:sufficient}

As we consider the transitions of the Büchi automaton as being labeled by
boolean expressions, it becomes more apparent that some pure IOs can be
redundant.  For instance, consider a Büchi automaton with
\(I = \{i\}, O = \{o_1, o_2\}\), but the only transitions compatible with \(i\) are
labeled \((i \land o_1)\) and \((i \land \neg o_1)\).  Pure IOs compatible with the first
label will be \((i \land o_1 \land o_2)\) and \((i \land o_1 \land \neg o_2)\), but
certainly, these two IOs have the same io-actions, and optimally, we would only
consider \((i \land o_1)\).  However, we should not consider \((i \land o_2)\), as it is
compatible with both transitions, but does not correspond to a pure IO.  We will
thus allow our main algorithm to select certain inputs and IOs:
\begin{definition}
  An IO (\resp input) is \emph{valid} if there is a pure IO (\resp input) with
  the same io-action (\resp input-action).  A set \(X\) of valid IOs is
  \emph{sufficient} if it represents all the possible io-actions of pure IOs:
  \(\{\ioact(x) \mid x \in X\} = \{\ioact(x) \mid x \text{ is a pure IO}\}.\)
  A sufficient set of inputs is defined similarly with input-actions.
\end{definition}

\subsection{Algorithm}\label{sec:algo}

We
solve \textbf{BackwardRealizability} by computing
\(\cpre^*\) explicitly:

\vspace{0.5em}

\begin{algorithm}[H]
  \LinesNumbered
\SetKwData{Downset}{\texttt{Downset}}
\SetKwFunction{Union}{Union}\SetKwFunction{FindCompress}{FindCompress}
\SetKwInput{Input}{Input}\SetKwInput{Output}{Output}

\Input{A Büchi automaton \cA, an integer \(k > 0\)}
\Output{Whether \((\exists \vec{v} \in  \cpre^*(\safe_k))[v_{q_0} \geq 0]\)}
\BlankLine

Possibly remove some useless states in \cA\;\label{alg:preproc}
Split states of \cA into boolean and nonboolean\;\label{alg:bool}
Let \Downset be a type for downsets using a vector type that possibly has a
boolean part\;\label{alg:types}
Let \(S = \safe_k\) of type \Downset\;
Compute a sufficient set \(E\) of  inputs\;\label{alg:inputs}
Compute the input-actions of \(E\)\;\label{alg:actions}
\While{true}{
  Pick an input-action \(a\) of \(E\)\;\label{alg:pick}
  \If{no action is returned}{\Return whether a vector in \(S\) maps \(q_0\) to a
    nonnegative value}
  \(S \leftarrow \cpre_a (S)\)\;
}
\caption{Main algorithm}\label{main_algo}
\end{algorithm}

\vspace{0.5em}

Our algorithm requires that the ``input-action picker'' used in line \ref{alg:pick}
decides whether we have reached a fixed point.  As the picker could check
whether \(S\) has changed, this is without loss of generality.

The computation of \(\cpre_a\) is the intuitive one, optimizations therein coming
from the internal representation of actions.  That is, it is implemented by
iterating through all io-actions compatible with \(a\), applying \(\bwd\) on
\(S\) for each of them, taking the union over all these applications, and finally
intersecting the result with \(S\).

\section{The many options at every line}\label{sec:implem}

\subsection{Preprocessing of the automaton (line \ref{alg:preproc})}\label{sec:preproc}

In this step, one can provide a heuristic that removes certain states that
do not contribute to the computation.  We provide an optional step that
detects \emph{surely losing states}, as presented in \cite{ggs14}.

\subsection{Boolean states (line \ref{alg:bool})}\label{sec:implem-bool}

We provide several implementations of the detection of boolean states, in
addition to an option to not detect them.  Our implementations are based on the
concept of \emph{bounded state}, as presented in \cite{bohy14}.  A state is
\emph{bounded} if it cannot be reached from a Büchi state that lies in a
nontrivial strongly connected component.  This 
can be detected in
several ways.

\subsection{Vectors and downsets (line \ref{alg:types})}\label{sec:vecds}

The most basic data structure in the main algorithm is that of a vector used to
give a value to the states.  We provide a handful of different vector
classes:
\begin{compactitem}
\item Standard C++ vector and array types (\cinline{std::vector},\linebreak
  \cinline{std::array}).  Note that arrays are of fixed size; our implementation
  precompiles arrays of different sizes (up to \(300\) by default),
  and defaults to vectors if more entries are needed.
\item Vectors and arrays backed by SIMD\footnote{SIMD: Single Instruction
  Multiple Data, a set of CPU instructions \& registers to compute
  component-wise operations on fixed-size vectors.} registers.  This makes use of the
  type \cinline{std::experimental::simd} and leverages modern CPU
  optimizations.
\end{compactitem}

Additionally, all these implementations can be glued to an array of booleans
(\cinline{std::bitset}) to provide a type that combines boolean and integer
values.  These types can optionally expose an integer that is compatible with
the partial order (here, the sum of all the elements in the vector: if
\(\vec{u}\) dominates \(\vec{v}\), then the sum of the elements in \(\vec{u}\) is
larger than that of \(\vec{v}\)).  This value can help the downset implementations
in sorting the vectors.

Downset types are built on top of a vector type.  We provide:
\begin{compactitem}
\item Implementations using sets or vectors of vectors, either containing only
  the dominating vectors, or containing explicitly all the vectors;
\item An implementation that relies on \(k\)-d trees, a space-partitioning data
  structure for organizing points in a \(k\)-dimensional space;
\item Implementations that store the vectors in specific bins depending on the
  information exposed by the vector type.
\end{compactitem}

\subsection{Selecting sufficient inputs (line \ref{alg:inputs})}\label{sec:suff}


Recall our discussion on sufficient inputs of Section~\ref{sec:sufficient}.  We
introduce the notion of \emph{terminal} IO following
the intuition that there
is
no restriction of the IO that would lead to a more specific action:
\begin{definition}
  An IO \(x\) is said to be \emph{terminal} if for every compatible IO \(y\), we
  have \(\ioact(x) \subseteq \ioact(y)\).  
  An \emph{input}~\(i\) is said to be
  \emph{terminal} if for every compatible input \(j\) we have
  \(\iact(i) \subseteq \iact(j)\).
\end{definition}

\begin{proposition}
  Any pure IO and any input is terminal.  Any terminal IO and any terminal input
  is valid.
\end{proposition}



Our approaches to input selection focus on efficiently searching for a
sufficient set of terminal IOs and inputs.
%
%
%
We present here a simple algorithm for computing a sufficient
set of terminal IOs.  

\begin{algorithm}
\SetKwInput{Input}{Input}\SetKwInput{Output}{Output}

\Input{A Büchi automaton \cA}
\Output{A sufficient set of terminal IOs}
\BlankLine
\(P \leftarrow \{\top\}\)\;
\For{every label \(x\) in the automaton}{
  \For{every element \(y\) in \(P\)}{
    \If{\(x \land y \neq \bot\)}{
      Delete \(y\) from \(P\)\;
      Insert \(x \land y\) in \(P\)\;
      \lIf{\(\neg x \land y \neq \bot\)}{
        insert \(\neg x \land y\) in \(P\)
      }
    }
  }
}
\Return \(P\)
\caption{Computing a sufficient set of terminal IOs}
\label{alg:sufficient}
\end{algorithm}
\noindent At this point, we provide 3 implementations of input selection:
\begin{compactitem}
\item No precomputation, i.e., return pure inputs/IOs;
\item Applying Algorithm~\ref{alg:sufficient} twice: for IOs and inputs;
\item Use a pure BDD approach to do the previous algorithm; this relies on extra
  variables to have the loop ``\textbf{for}\emph{ every element \(y\) in \(P\)}''
  iterate \emph{only} over elements \(y\) that satisfy \(x \land y \neq \bot\).
\end{compactitem}

\subsection{Precomputing actions (line \ref{alg:actions})}

Since computing \(\cpre_i\) for an input \(i\) requires to go through
\(\iact(i)\), possibly going back to the automaton and iterating through all
transitions, it may be beneficial to precompute this set.  We provide this step
as an optional optimization that is intertwined with the computation of a
sufficient set of IOs; for instance, rather than iterating through labels in
Algorithm~\ref{alg:sufficient}, one could iterate through all transitions, and
store the set of transitions that are compatible with each terminal IO on the
fly.

\subsection{Main loop: Picking input-actions (line \ref{alg:pick})}\label{sec:io}

We provide several implementations of the input-action picker:
\begin{compactitem}
\item Return each input-action in turn, until no change has occurred to \(S\)
  while going through all possible input-actions;
\item Search for an input-action that is certain to change \(S\).  This is based
  on the concept of \emph{critical input} as presented in~\cite{bohy14}.  This
  is reliant on how input-actions are ordered themselves, so we provide multiple
  options (using a priority queue to prefer inputs that were recently returned,
  randomize part of the array of input-actions, and randomize the whole array).
\end{compactitem}

\section{Checking nonrealizability}\label{sec:unreal}

As mentioned in the preliminaries, for large values of $k$ the 
\textbf{BackwardRealizability} problem is equivalent to a non-zero sum game
whose payoff set is the complement of the language of the given automaton.
More precisely, for small values of $k$, a negative answer for the
\textbf{BackwardRealizability} problem does not imply the output player does
not win the game. Instead, if one is interested in whether the output player
wins, a property known as determinacy~\cite{borel} can be leveraged to instead
ask whether a complementary property holds: does the input player win the
game?

We thus need to build an automaton \(\mathcal{B}\) for which a positive answer to
the \textbf{BackwardRealizability} translates to the previous property.  To do
so, we can consider the negation of the input formula, \(\neg\phi\), and inverse the
roles of the players, that is, swap the inputs and outputs.  However, to make
sure the semantics of the game is preserved, we also need to have the input
player play first, and the output player \emph{react} to the input player's
move.  To do so, we simply need to have the outputs moved \emph{one step
  forward} (in the future, in the LTL sense).  This can be done directly on the
input formula, by putting an \(X\) (neXt) operator on each output.  This can
however make the formula much more complex.

We propose an alternative to this: Obtain the automaton for \(\neg \phi\), then push the
outputs one state forward.  This means that a transition \((p, \langle i, o \rangle, q)\) is
translated to a transition \((p, i, q)\), and the output \(o\) should be fired from
\(q\).  In practice, we would need to remember that output, and this would require
the construction to consider every state \((q, o)\), augmenting the number of
states tremendously.  Algorithm~\ref{alg:caddys-madness} for this task, however,
tries to minimize the number of states \((q, o)\) necessary by considering nonpure
outputs that maximally correspond to a nonpure input compatible with the
original transition label.

\vspace{1em}

\begin{algorithm}[H]
\SetKwInput{Input}{Input}\SetKwInput{Output}{Output}

\Input{A Büchi automaton \cA with initial state $q_0$}
\Output{The states $S$ and transitions $\Delta$ of the B\"uchi automaton
$\mathcal{B}$}
\BlankLine
\(S,V \leftarrow \{(q_0,\top)\}\), \(\Delta \leftarrow \{\}\)\;
\While{\(V\) is nonempty}{
  Pop $(p,o)$ from $V$\;
  \For{every \((p,x,q) \in \delta\)}{
    $y \leftarrow x$\;
    \While{$y \neq \bot$}{
      Let $o'$ be a pure output compatible with $y$\;
      Let $i$ be an input s.t. $i \land o' \equiv y \land o'$\;
      $o'' \leftarrow \exists I.\. i \land y$\;
      Add $(\langle p, o\rangle, o \land i, \langle q, o''\rangle)$ to
      $\Delta$\;
      If $(q,o')$ is not in $S$, add it to $S$ and $V$\;
      $y \leftarrow y \land \lnot i$\;
    }
  }
}
\Return \(S,\Delta\)
\caption{Modifying \cA so that the outputs are shifted forward}
\label{alg:caddys-madness}
\end{algorithm}

\section{Benchmarks}\label{sec:benchmarks}

\subsection{Protocol}

For the past few years, the yardstick of performance for synthesis tools is the
SYNTCOMP competition~\cite{syntcomp17}.  The organizers provide a bank of nearly a
thousand LTL formulas, and candidate tools are ran with a time limit of one hour
on each of them.  The tool that solves the most instances in this timeframe wins
the competition.

To benchmark our tool, we selected all the LTL formulas that were accepted in
less than 100 seconds by \emph{any} tool that competed in the 2021 SYNTCOMP
competition.  These are 879 out of 945 tests.  Notably, 864 of these tests were
solved in less than 20 seconds by some tool, and among the 66 tests left out, 50
were not solved by any tool.  This displays a usual trend of synthesis tools:
either they solve an instance fast, or they are unlikely to solve it at all.  To
better focus on the fine performance differences between the tools, we set a
timeout of 15 seconds for all tests.

We compared Acacia-Bonsai against itself using different choices of options, and
against Acacia+~\cite{bbfjr12}, Strix~\cite{msl18}, and
ltlsynt~\cite{duret.16.atva2}.  The benchmarks were completed on a headless
Linux computer with the following specifications:
\begin{compactitem}
\item CPU: Intel\textregistered{} Core\texttrademark{} i7-8700 CPU @ 3.20GHz.  This CPU
  has 6 hyperthreaded cores, meaning that 12 threads can run concurrently.  It
  supports Intel\textregistered{} AVX2, meaning that it has SIMD registers of up to
  256 bits.
\item Memory: The CPU has 12 MiB of cache, the computer has 16 GiB of DDR4-2666
  RAM.
\end{compactitem}

\subsection{Results}

\paragraph{The options of Acacia-Bonsai.}  We compared about 30 different
configurations of Acacia-Bonsai, in order to single out the best combination of
options.  
\begin{compactitem}
\item Preprocessing of the automaton (\Cref{sec:preproc}). This seems to have
  little impact, although a handful of tests saw an important boost.  Overall,
  the performances were \emph{worse} with automaton preprocessing, owing to the
  cost of computing the surely loosing states.  Overall, we elected to leave the
  option activated in our best configuration.
\item Boolean states (\Cref{sec:implem-bool}). Using boolean states boosted
  performances by a marginal amount when SIMD was activated, but had a more
  important impact when SIMD was deactivated.  This follows the intuition that,
  thanks to bitmasks, boolean states allow for vectorized computing, and they
  are thus of a lesser impact when native vectorized computing is possible.
\item Vectors and downsets (\Cref{sec:vecds}). For the vector implementation to
  become a bottleneck, specific implementations of downsets have to be selected.
  Although downsets implemented using \(k\)-d trees do not outperform the other
  implementations with SIMD deactivated, they perform significantly better with
  SIMD.  We show in our main graphic the impact of deactivating SIMD with
  \(k\)-d trees.
\item Precomputing a sufficient set of inputs and IO (\Cref{sec:suff}).  This
  comes with a significant boost in speed (shown in the graph below when
  comparing all implementations).  Among our different implementations to find a
  sufficient set, \Cref{alg:sufficient} turned out to offer the best
  performances.
\item Picking input-actions (\Cref{sec:io}).  The approaches performed
  equivalently, with a slight edge for the choice of critical inputs without
  randomizing or priority queue.
\item Nonrealizability (\Cref{sec:unreal}). Computation of the nonrealizability
  automaton using \Cref{alg:caddys-madness}, rather than applying \(X\) on the
  outputs of the formula, allowed to solve 14\% more instances.  In practice, we
  elected to create two processes for nonrealizability, one for each option,
  allowing for parallel computation on the two resulting automata.
\end{compactitem}

\paragraph{Acacia-Bonsai and foes.}  The following graph shows the performance
of the tools.  Since these tools tend to solve a lot of instances under one
second, we elected to present this graphic with a logarithmic y-axis.  Thanks to
this, the cost of entry of SIMD instructions is also emphasized.

\vspace{1em}

\noindent \includegraphics[width=\textwidth]{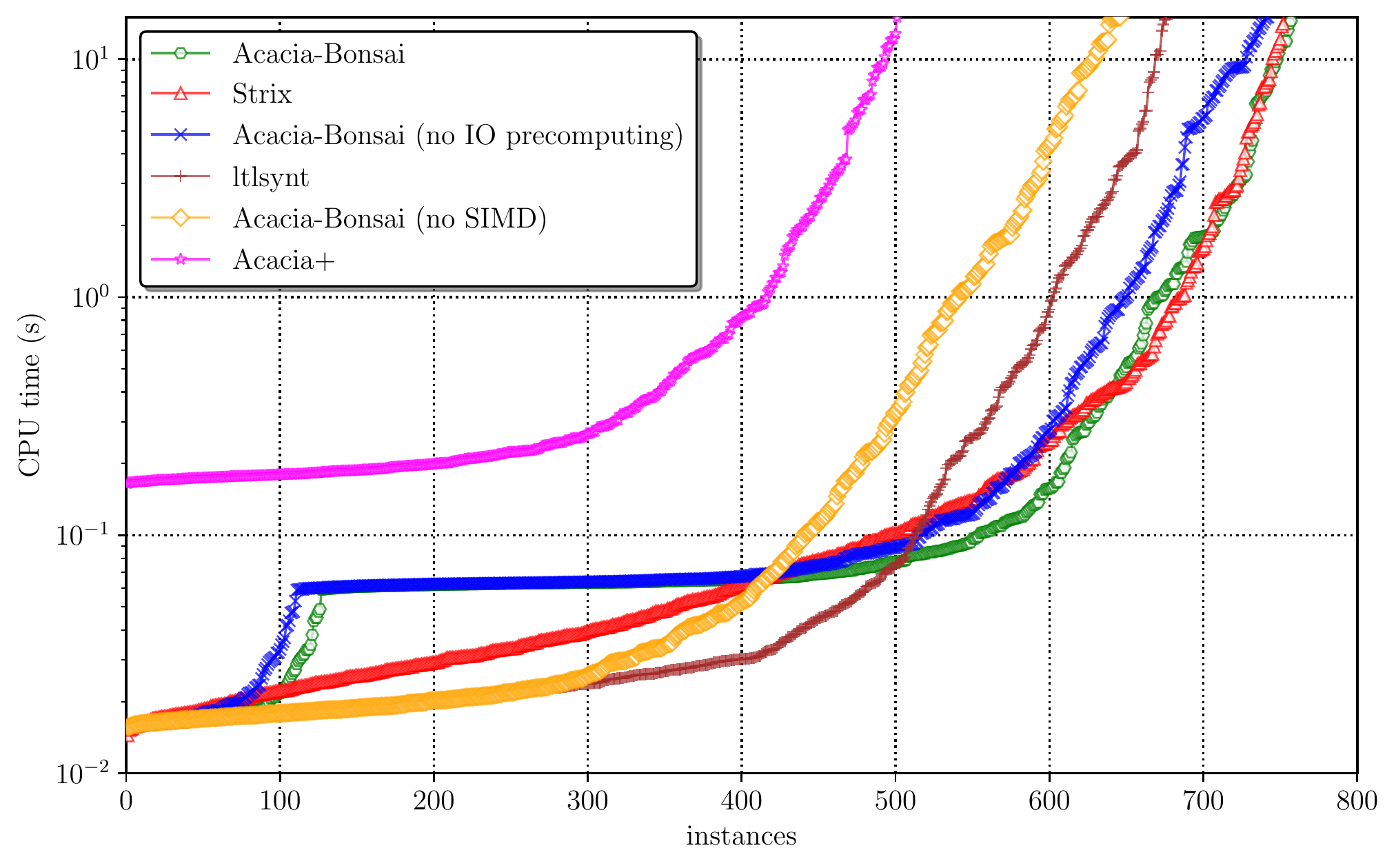}


\section{Conclusion}

We provided six degrees of freedom in the main algorithm for downset-based LTL
realizability and implemented multiple options for each of these degrees.  In
this paper, we presented the main ideas behind these.  Experiments show that
this careful reimplementation surpasses the performances of the original
Acacia+, making Acacia-Bonsai competitive against modern LTL realizability
tools. Our implementation can be found at \url{https://github.com/gaperez64/acacia-bonsai/}.

\bibliographystyle{splncs04}
\bibliography{refs}

\begin{thebibliography}{10}
\providecommand{\url}[1]{\texttt{#1}}
\providecommand{\urlprefix}{URL }
\providecommand{\doi}[1]{https://doi.org/#1}

\bibitem{abbdfhinprs17}
Ahmed, Z., Benqu{\'{e}}, D., Berezin, S., Dahl, A.C.E., Fisher, J., Hall, B.A.,
  Ishtiaq, S., Nanavati, J., Piterman, N., Riechert, M., Skoblov, N.: Bringing
  {LTL} model checking to biologists. In: Bouajjani, A., Monniaux, D. (eds.)
  Verification, Model Checking, and Abstract Interpretation - 18th
  International Conference, {VMCAI} 2017, Paris, France, January 15-17, 2017,
  Proceedings. Lecture Notes in Computer Science, vol. 10145, pp. 1--13.
  Springer (2017). \doi{10.1007/978-3-319-52234-0\_1},
  \url{https://doi.org/10.1007/978-3-319-52234-0\_1}

\bibitem{bltv20}
Bansal, S., Li, Y., Tabajara, L.M., Vardi, M.Y.: Hybrid compositional reasoning
  for reactive synthesis from finite-horizon specifications. In: The
  Thirty-Fourth {AAAI} Conference on Artificial Intelligence, {AAAI} 2020, The
  Thirty-Second Innovative Applications of Artificial Intelligence Conference,
  {IAAI} 2020, The Tenth {AAAI} Symposium on Educational Advances in Artificial
  Intelligence, {EAAI} 2020, New York, NY, USA, February 7-12, 2020. pp.
  9766--9774. {AAAI} Press (2020),
  \url{https://aaai.org/ojs/index.php/AAAI/article/view/6528}

\bibitem{bohy14}
Bohy, A.: Antichain based algorithms for the synthesis of reactive systems.
  Ph.D. thesis, University of Mons (2014)

\bibitem{bbfjr12}
Bohy, A., Bruy{\`{e}}re, V., Filiot, E., Jin, N., Raskin, J.: Acacia+, a tool
  for {LTL} synthesis. In: Madhusudan, P., Seshia, S.A. (eds.) CAV. LNCS,
  vol.~7358, pp. 652--657. Springer (2012).
  \doi{10.1007/978-3-642-31424-7\_45},
  \url{https://doi.org/10.1007/978-3-642-31424-7\_45}

\bibitem{bvpyb16}
Bombara, G., Vasile, C.I., Penedo, F., Yasuoka, H., Belta, C.: A decision tree
  approach to data classification using signal temporal logic. In: Abate, A.,
  Fainekos, G.E. (eds.) Proceedings of the 19th International Conference on
  Hybrid Systems: Computation and Control, {HSCC} 2016, Vienna, Austria, April
  12-14, 2016. pp. 1--10. {ACM} (2016). \doi{10.1145/2883817.2883843},
  \url{https://doi.org/10.1145/2883817.2883843}

\bibitem{cm19}
Camacho, A., McIlraith, S.A.: Learning interpretable models expressed in linear
  temporal logic. In: Benton, J., Lipovetzky, N., Onaindia, E., Smith, D.E.,
  Srivastava, S. (eds.) Proceedings of the Twenty-Ninth International
  Conference on Automated Planning and Scheduling, {ICAPS} 2018, Berkeley, CA,
  USA, July 11-15, 2019. pp. 621--630. {AAAI} Press (2019),
  \url{https://aaai.org/ojs/index.php/ICAPS/article/view/3529}

\bibitem{duret.16.atva2}
Duret-Lutz, A., Lewkowicz, A., Fauchille, A., Michaud, T., Renault, E., Xu, L.:
  Spot 2.0 --- a framework for {LTL} and $\omega$-automata manipulation. In:
  Proceedings of the 14th International Symposium on Automated Technology for
  Verification and Analysis (ATVA'16). Lecture Notes in Computer Science,
  vol.~9938, pp. 122--129. Springer (Oct 2016).
  \doi{10.1007/978-3-319-46520-3_8}

\bibitem{ekrs17}
Esparza, J., Kret{\'{\i}}nsk{\'{y}}, J., Raskin, J., Sickert, S.: From {LTL}
  and limit-deterministic {B}{\"{u}}chi automata to deterministic parity
  automata. In: Legay, A., Margaria, T. (eds.) Tools and Algorithms for the
  Construction and Analysis of Systems - 23rd International Conference, {TACAS}
  2017, Held as Part of the European Joint Conferences on Theory and Practice
  of Software, {ETAPS} 2017, Uppsala, Sweden, April 22-29, 2017, Proceedings,
  Part {I}. Lecture Notes in Computer Science, vol. 10205, pp. 426--442 (2017).
  \doi{10.1007/978-3-662-54577-5\_25},
  \url{https://doi.org/10.1007/978-3-662-54577-5\_25}

\bibitem{eks16}
Esparza, J., Kret{\'{\i}}nsk{\'{y}}, J., Sickert, S.: From {LTL} to
  deterministic automata - {A} {S}afraless compositional approach. Formal
  Methods Syst. Des.  \textbf{49}(3),  219--271 (2016).
  \doi{10.1007/s10703-016-0259-2},
  \url{https://doi.org/10.1007/s10703-016-0259-2}

\bibitem{ffrt17}
Faymonville, P., Finkbeiner, B., Rabe, M.N., Tentrup, L.: Encodings of bounded
  synthesis. In: Legay, A., Margaria, T. (eds.) TACAS. LNCS, vol. 10205, pp.
  354--370 (2017). \doi{10.1007/978-3-662-54577-5\_20},
  \url{https://doi.org/10.1007/978-3-662-54577-5\_20}

\bibitem{fjr09}
Filiot, E., Jin, N., Raskin, J.: An antichain algorithm for {LTL}
  realizability. In: Bouajjani, A., Maler, O. (eds.) Computer Aided
  Verification, 21st International Conference, {CAV} 2009, Grenoble, France,
  June 26 - July 2, 2009. Proceedings. Lecture Notes in Computer Science,
  vol.~5643, pp. 263--277. Springer (2009).
  \doi{10.1007/978-3-642-02658-4\_22},
  \url{https://doi.org/10.1007/978-3-642-02658-4\_22}

\bibitem{ggs14}
Geeraerts, G., Goossens, J., Stainer, A.: Synthesising succinct strategies in
  safety and reachability games. In: Ouaknine, J., Potapov, I., Worrell, J.
  (eds.) RP. LNCS, vol.~8762, pp. 98--111. Springer (2014).
  \doi{10.1007/978-3-319-11439-2\_8},
  \url{https://doi.org/10.1007/978-3-319-11439-2\_8}

\bibitem{gv16}
Giacomo, G.D., Vardi, M.Y.: {LTL}\({}_{\mbox{f}}\) and {LDL}\({}_{\mbox{f}}\)
  synthesis under partial observability. In: Kambhampati, S. (ed.) Proceedings
  of the Twenty-Fifth International Joint Conference on Artificial
  Intelligence, {IJCAI} 2016, New York, NY, USA, 9-15 July 2016. pp.
  1044--1050. {IJCAI/AAAI} Press (2016),
  \url{http://www.ijcai.org/Abstract/16/152}

\bibitem{gnpw20}
Gutierrez, J., Najib, M., Perelli, G., Wooldridge, M.J.: Automated temporal
  equilibrium analysis: Verification and synthesis of multi-player games.
  Artif. Intell.  \textbf{287},  103353 (2020).
  \doi{10.1016/j.artint.2020.103353},
  \url{https://doi.org/10.1016/j.artint.2020.103353}

\bibitem{hp84}
Harel, D., Pnueli, A.: On the development of reactive systems. In: Apt, K.R.
  (ed.) Logics and Models of Concurrent Systems - Conference proceedings,
  Colle-sur-Loup (near Nice), France, 8-19 October 1984. {NATO} {ASI} Series,
  vol.~13, pp. 477--498. Springer (1984). \doi{10.1007/978-3-642-82453-1\_17},
  \url{https://doi.org/10.1007/978-3-642-82453-1\_17}

\bibitem{syntcomp17}
Jacobs, S., Basset, N., Bloem, R., Brenguier, R., Colange, M., Faymonville, P.,
  Finkbeiner, B., Khalimov, A., Klein, F., Michaud, T., P{\'{e}}rez, G.A.,
  Raskin, J., Sankur, O., Tentrup, L.: The 4th reactive synthesis competition
  {(SYNTCOMP} 2017): Benchmarks, participants {\&} results. In: Fisman, D.,
  Jacobs, S. (eds.) Proceedings Sixth Workshop on Synthesis, SYNT@CAV 2017,
  Heidelberg, Germany, 22nd July 2017. {EPTCS}, vol.~260, pp. 116--143 (2017).
  \doi{10.4204/EPTCS.260.10}, \url{https://doi.org/10.4204/EPTCS.260.10}

\bibitem{kpv06}
Kupferman, O., Piterman, N., Vardi, M.Y.: Safraless compositional synthesis.
  In: Ball, T., Jones, R.B. (eds.) Computer Aided Verification, 18th
  International Conference, {CAV} 2006, Seattle, WA, USA, August 17-20, 2006,
  Proceedings. Lecture Notes in Computer Science, vol.~4144, pp. 31--44.
  Springer (2006). \doi{10.1007/11817963\_6},
  \url{https://doi.org/10.1007/11817963\_6}

\bibitem{lpb15}
Lemieux, C., Park, D., Beschastnikh, I.: General {LTL} specification mining
  {(T)}. In: Cohen, M.B., Grunske, L., Whalen, M. (eds.) 30th {IEEE/ACM}
  International Conference on Automated Software Engineering, {ASE} 2015,
  Lincoln, NE, USA, November 9-13, 2015. pp. 81--92. {IEEE} Computer Society
  (2015). \doi{10.1109/ASE.2015.71}, \url{https://doi.org/10.1109/ASE.2015.71}

\bibitem{lms20}
Luttenberger, M., Meyer, P.J., Sickert, S.: Practical synthesis of reactive
  systems from {LTL} specifications via parity games. Acta Informatica
  \textbf{57}(1-2),  3--36 (2020). \doi{10.1007/s00236-019-00349-3},
  \url{https://doi.org/10.1007/s00236-019-00349-3}

\bibitem{borel}
Martin, D.A.: Borel determinacy. Annals of Mathematics  \textbf{102}(2),
  363--371 (1975), \url{http://www.jstor.org/stable/1971035}

\bibitem{msl18}
Meyer, P.J., Sickert, S., Luttenberger, M.: Strix: Explicit reactive synthesis
  strikes back! In: Chockler, H., Weissenbacher, G. (eds.) Computer Aided
  Verification - 30th International Conference, {CAV} 2018, Held as Part of the
  Federated Logic Conference, FloC 2018, Oxford, UK, July 14-17, 2018,
  Proceedings, Part {I}. Lecture Notes in Computer Science, vol. 10981, pp.
  578--586. Springer (2018). \doi{10.1007/978-3-319-96145-3\_31},
  \url{https://doi.org/10.1007/978-3-319-96145-3\_31}

\bibitem{piterman07}
Piterman, N.: From nondeterministic {B}{\"{u}}chi and {S}treett automata to
  deterministic parity automata. Log. Methods Comput. Sci.  \textbf{3}(3)
  (2007). \doi{10.2168/LMCS-3(3:5)2007},
  \url{https://doi.org/10.2168/LMCS-3(3:5)2007}

\bibitem{pnueli77}
Pnueli, A.: The temporal logic of programs. In: 18th Annual Symposium on
  Foundations of Computer Science, Providence, Rhode Island, USA, 31 October -
  1 November 1977. pp. 46--57. {IEEE} Computer Society (1977).
  \doi{10.1109/SFCS.1977.32}, \url{https://doi.org/10.1109/SFCS.1977.32}

\bibitem{pr89}
Pnueli, A., Rosner, R.: On the synthesis of a reactive module. In: Conference
  Record of the Sixteenth Annual {ACM} Symposium on Principles of Programming
  Languages, Austin, Texas, USA, January 11-13, 1989. pp. 179--190. {ACM} Press
  (1989). \doi{10.1145/75277.75293}, \url{https://doi.org/10.1145/75277.75293}

\bibitem{safra88}
Safra, S.: On the complexity of omega-automata. In: 29th Annual Symposium on
  Foundations of Computer Science, White Plains, New York, USA, 24-26 October
  1988. pp. 319--327. {IEEE} Computer Society (1988).
  \doi{10.1109/SFCS.1988.21948}, \url{https://doi.org/10.1109/SFCS.1988.21948}

\bibitem{apsec20}
Shi, Y., Xiao, S., Li, J., Guo, J., Pu, G.: Sat-based automata construction for
  {LTL} over finite traces. In: 27th Asia-Pacific Software Engineering
  Conference, {APSEC} 2020, Singapore, December 1-4, 2020. pp. 1--10. {IEEE}
  (2020). \doi{10.1109/APSEC51365.2020.00008},
  \url{https://doi.org/10.1109/APSEC51365.2020.00008}

\bibitem{tushy17}
Tomita, T., Ueno, A., Shimakawa, M., Hagihara, S., Yonezaki, N.: Safraless
  {LTL} synthesis considering maximal realizability. Acta Informatica
  \textbf{54}(7),  655--692 (2017). \doi{10.1007/s00236-016-0280-3},
  \url{https://doi.org/10.1007/s00236-016-0280-3}

\bibitem{vw84}
Vardi, M.Y., Wolper, P.: Automata theoretic techniques for modal logics of
  programs (extended abstract). In: DeMillo, R.A. (ed.) Proceedings of the 16th
  Annual {ACM} Symposium on Theory of Computing, April 30 - May 2, 1984,
  Washington, DC, {USA}. pp. 446--456. {ACM} (1984).
  \doi{10.1145/800057.808711}, \url{https://doi.org/10.1145/800057.808711}

\bibitem{ztlpv17}
Zhu, S., Tabajara, L.M., Li, J., Pu, G., Vardi, M.Y.: A symbolic approach to
  safety {LTL} synthesis. In: Strichman, O., Tzoref{-}Brill, R. (eds.) HVC.
  LNCS, vol. 10629, pp. 147--162. Springer (2017).
  \doi{10.1007/978-3-319-70389-3\_10},
  \url{https://doi.org/10.1007/978-3-319-70389-3\_10}

\end{thebibliography}

\end{document}